\documentclass[12pt,preprint]{aastex}

\slugcomment{To appear in AJ., XX.}
 
\shorttitle{SNLS SN\,Ia in Galaxy Clusters}
\shortauthors{Graham et al.}

\begin{document}
 
\title{Type Ia Supernovae Rates and Galaxy Clustering from the CFHT Supernova Legacy Survey}

\author{M.L. Graham\altaffilmark{1}, C.J. Pritchet\altaffilmark{1}, M. Sullivan\altaffilmark{2}, S.D.J. Gwyn\altaffilmark{1,3}, J.D. Neill\altaffilmark{1,4}, E.Y. Hsiao\altaffilmark{1}, P. Astier\altaffilmark{5}, D. Balam\altaffilmark{1}, C. Balland\altaffilmark{5}, S. Basa\altaffilmark{6}, R.G. Carlberg\altaffilmark{2}, A. Conley\altaffilmark{2}, D. Fouchez\altaffilmark{7}, J. Guy\altaffilmark{5}, D. Hardin\altaffilmark{5}, I.M. Hook\altaffilmark{8}, D.A. Howell\altaffilmark{2}, R. Pain\altaffilmark{5}, K. Perrett\altaffilmark{2}, N. Regnault\altaffilmark{5}, S. Baumont\altaffilmark{5}, J. Le Du\altaffilmark{7}, C. Lidman\altaffilmark{9}, S. Perlmutter\altaffilmark{10}, P. Ripoche\altaffilmark{7}, N. Suzuki\altaffilmark{10}, E.S. Walker\altaffilmark{8}, T. Zhang\altaffilmark{6}}

\altaffiltext{1}{Department of Physics and Astronomy, University of Victoria, PO Box 3055 STN CSC, Victoria BC V8T 1M8, Canada}
\altaffiltext{2}{Department of Astronomy and Astrophysics, University of Toronto, 60 St. George Street, Toronto ON M5S 3H8, Canada}
\altaffiltext{3}{Canadian Astronomy Data Centre, NRC Herzberg Institute for Astrophysics, 5071 West Saanich Road, Victoria BC V9E 2E7, Canada}
\altaffiltext{4}{California Institute of Technology, E. California Blvd., Pasadena CA 91125, USA}
\altaffiltext{5}{LPNHE, CNRS-IN2P3 and Universit\'{e}s Paris VI \& VII, 4 place Jussieu, 75252 Paris Cedex 05, France}
\altaffiltext{6}{LAM, CNRS, BP8, Traverse du Siphon, 13376 Marseille Cedex 12, France}
\altaffiltext{7}{CPPM, CNRS-IN2P3 and Universit\'{e} Aix-Marseille II, Case 907, 13288 Marseille Cedex 9, France}
\altaffiltext{8}{University of Oxford Astrophysics, Denys Wilkinson Building, Keble Road, Oxford OX1 3RH, UK}
\altaffiltext{9}{ESO, Alonso de Cordova 3107, Vitacura, Casilla 19001, Santiago 19, Chile}
\altaffiltext{10}{Lawrence Berkeley National Laboratory, 1 Cyclotron Road, Berkeley CA 94720 USA}

%%%%%%%%%%%%%%%%%%%%%%%%%%%%%%%%%%%%%%%%%%%%%%%%%%%%%%%%%%%%%%%%%%%%%%%%%%%%%%%%%%%%%%%%%%%%%%%%%%%%%%%%%%%%%%%%%%%%%%%
%\clearpage
\begin{abstract}
The Canada-France-Hawaii Telescope Supernova Legacy Survey (SNLS) has created a
large homogeneous database of intermediate redshift ($0.2 < z < 1.0$) type Ia
supernovae (SNe\,Ia). The SNLS team has shown that correlations exist between
SN Ia rates, properties, and host galaxy star formation rates.
The SNLS SN\,Ia database has now been combined with a photometric redshift galaxy
catalog and an optical galaxy cluster catalog to investigate the possible influence
of galaxy clustering on the SN\,Ia rate, over and above the expected effect due to
the dependence of SFR on clustering through the morphology-density relation.
We identify three cluster SNe\,Ia, plus three additional possible cluster SNe\,Ia,
and find the SN\,Ia rate per unit mass in clusters at intermediate redshifts is
consistent with the rate per unit mass in field early-type galaxies and the
SN\,Ia cluster rate from low redshift cluster targeted surveys.
We also find the number of SNe\,Ia in cluster environments
to be within a factor of two of expectations from the two component SN\,Ia rate model.
\end{abstract}
 
\keywords{supernovae:general --- galaxies:clusters:general}

%%%%%%%%%%%%%%%%%%%%%%%%%%%%%%%%%%%%%%%%%%%%%%%%%%%%%%%%%%%%%%%%%%%%%%%%%%%%%%%%%%%%%%%%%%%%%%%%%%%%%%%%%%%%%%%%%%%%%%%
 
\section{INTRODUCTION}
\label{S1Intro}
 
The importance of understanding Type Ia supernovae (SNe\,Ia)
has risen greatly in recent years because of their pivotal role
in demonstrating the accelerated expansion of the universe \citep{R98,P99,A06,W-V07}.
SNe\,Ia are generally agreed to be carbon-oxygen 
white dwarfs which have accreted sufficient mass from their
companion star to initiate a thermonuclear explosion. Despite this consensus, several models exist  
for the companion type, the accretion process, the delay time 
between star formation and SN\,Ia explosion, 
and the explosion mechanism \citep{Hill00,Hoflich03}.
Photometric and spectroscopic observations, and 
explosion simulations, are making inroads towards a 
coherent picture of SN\,Ia events \citep{Mazzali07}.
Another (albeit indirect) path of investigation
is studying the SN\,Ia rate in different stellar populations.

Compilation of the Cappellaro et al. (1999) supernova catalog with infrared data 
% from the 2 Micron All Sky Survey (2MASS) Extended Source Catalog
led Mannucci et al. (2005) to conclude that SN\,II, SN\,Ib/c, and SN\,Ia rates per unit stellar mass
are directly correlated with host morphology and (B-K) colour, and to infer a correlation
with host star formation activity. 
Based on this, Scannapieco \& Bildsten (2005) expressed the SN\,Ia rate per unit stellar mass as the
sum of a ``delayed'' component from old and a ``prompt'' component from young stellar populations.
They parametrized the two components as A and B, proportional to the mass and star formation rate (SFR)
of a galaxy respectively, and also found the ``prompt'' component to account for the the discrepancy
between low observed rates of cluster SNe\,Ia and the high cluster iron abundance \citep{Maoz04}.
This ``A+B'' two component model has since been confirmed at intermediate redshifts with the large
SNLS catalog \citep{S06a}; the observations can also be matched with progenitor populations that have
distributions of delay times as described by Mannucci et al. (2006) and Pritchet et al. (2007).

From the two component model, the SN\,Ia rate in galaxy clusters 
is expected to be lower than in the field due to the morphology-density relation
\citep{Postman84}: cluster galaxies are predominantly of early-type
with little or no star formation.
However, since clusters contain only a small fraction of the stellar mass
of the Universe, it is conceivable that some hitherto undetected
influence in such exotic environments could affect the SN\,Ia rate. For 
example, the fraction of binary stars, or the rate of mass accretion 
onto the white dwarf, could be enhanced. The recent discovery of an 
enhanced nova rate in the core of elliptical galaxy M87 is evidence 
for the latter \citep{M07}. A second example is the detected SN\,Ia rate
enhancement in radio-loud early-type galaxies \citep{DV05}; such galaxies
tend to be very luminous ellipticals in the centers of large clusters.
 
The Wise Observatory Optical Transient Survey (WOOTS)
targeting 140 Abell clusters confirmed the SNe\,Ia rate per
unit mass in low redshift galaxy clusters to be consistent with the rate in 
early-type galaxies \citep{KS07}. Recently, Mannucci et al. (2007) analyzed
the Cappellaro et al. (1999) sample of 136 low redshift SNe ($\rm z<0.04$)
and found the SN\,Ia rate in cluster early-type galaxies is enhanced by a
factor of $\gtrsim 3$ over field early-type galaxies. The large database of intermediate to high 
redshift SNe\,Ia compiled by the Canada-France Hawaii Telescope
Supernova Legacy Survey (CFHT SNLS) \citep{A06}, and the publicly available photometric
redshift galaxy and cluster catalogs for SNLS fields \citep{I06,LFO} 
are ideal for extending these investigations to higher redshifts. As SNLS
is not a cluster-targeted survey, further advantages over Sharon et al. (2007) include
using a flexible parametrization of galaxy clustering in the environments of SNe\,Ia,
and determining simultaneous field SN\,Ia rates for comparison to cluster rates.

\S~\ref{S2Obs} describes the SN\,Ia, galaxy, and cluster 
catalogs used in this experiment. \S~\ref{S3SigParam}
and \S~\ref{S4LFO} present two independent approaches to statistically 
evaluate the effects of clustering on the SN\,Ia
rate per unit mass. \S~\ref{S3SigParam} uses a continuous clustering strength
estimator to compare the SN\,Ia rate per unit mass inside
and outside clustered environments. \S~\ref{S4LFO} identifies six
SN\,Ia in Olsen et al. (2007) galaxy clusters, considers
the probability of these observations based on expectations from 
the two component rate model, and calculates the SN\,Ia rate per
unit mass in clusters. \S~\ref{S5Conc} reviews and discusses
the papers main findings.

A flat cosmology with $\rm H_0=70 \ km \ s^{-1} \ Mpc^{-1}$, $\Omega_{Lambda}=0.7$, and $\Omega_M=0.3$ is used.

%%%%%%%%%%%%%%%%%%%%%%%%%%%%%%%%%%%%%%%%%%%%%%%%%%%%%%%%%%%%%%%%%%%%%%%%%%%%%%%%%%%%%%%%%%%%%%%%%%%%%%%%%%%%%%%%%%%%%%%
 
\section{OBSERVATIONS}
\label{S2Obs}
 
Three catalogs are described here, all of which were generated from the
four Deep fields of the Canada-France-Hawaii Telescope Legacy Survey:
the Supernova Legacy Survey catalog of type Ia
supernovae\footnote{http://legacy.astro.utoronto.ca}, the Ilbert et al. (2006)
photometric redshift galaxy catalog, and the Olsen et al. (2007) optical
galaxy cluster catalog.

\subsection{The CFHT Supernova Legacy Survey Catalog}
\label{sS2SNLS}
 
SNLS images four 1 deg$^2$ Deep fields (D1--D4) every three to four nights (when
visible) in four MegaCam filters ($g_M,r_M,i_M,z_M$) to a depth $i_M \simeq
25$. At the end of its five year program, the SNLS will have discovered $\sim500$ type Ia
supernova with its imaging and spectroscopic programs \citep{H05},
and be a valuable contributor to collaborative efforts in constraining
the dark energy equation of state parameter $w$ \citep{A06,Sper07}. For this 
project we use the 343 SNe\,Ia spectroscopically identified prior to 2007 September 29.

\subsection{Ilbert Photometric Redshift Galaxy Catalog}
\label{sS2IlbCat}
 
The Ilbert et al. (2006) galaxy photometric redshift catalog\footnote{http://terapix.iap.fr} 
covers the four SNLS Deep fields. They incorporate
VIMOS VLT Deep Survey spectroscopic redshifts to calibrate the spectral energy
distribution (SED) fitting routine, resulting in photometric redshifts
with an accuracy of $\sigma_{\Delta z / (1+z)} =0.029$ for $i_{AB}<24$,
and a fraction of catastrophic errors of 1\% at $17.5<i_{AB}<21.5$,
increasing to 10\% at $23.5<i_{AB}<24$ \citep{I06}.
In optimizing the photo-z calculation, the accuracies of the SEDs types are
compromised (Ilbert, private communication), and the distribution of 
SED types is discontinuous.  To solve this problem, we use 51 SEDs 
interpolated from Coleman et al. (1980) and Kinney et al. (1996) templates\footnote{http://www.astro.uvic.ca/$\sim$gwyn/cfhtls},
and fit them to the apparent magnitudes and
photometric redshifts of the catalog galaxies. We then
calculate $\ensuremath{K}$ corrections and absolute magnitudes, and
estimate galaxy stellar masses and star formation rates using fits of
this library of SEDs to the models of Buzzoni et al. (2005), which we
find to agree well (within a factor of ~2) with those derived from PEGASE models \citep{S06b}.

To ensure catalog purity, we restrict the catalog as recommended 
by Ilbert et al. (2006). For example, galaxies must be detected in $i_M$,
the flag values are used to mask galaxies near foreground stars,
the number of bands used for redshift fit must be at least 3,
and galaxies with a second peak in their redshift Probability Distribution Function
indicate a likely catastrophic failure, thus are excluded \citep{I06}.
In addition we impose the limits $i_M < 25$ and $z < 1.2$, and that the ``object'' parameter
must be equal to zero indicating the object is a galaxy.

\subsection{Olsen Optical Galaxy Cluster Catalog}
\label{sS2LFO}
 
The Olsen et al. (2007) cluster catalog derived from the Terapix data release in
August 2005 (T0002) includes photometric redshifts and an optical grade (A--D).
Olsen clusters are restricted to those of grade A, meaning a concentration
of similarly colored galaxies is visible, resulting in 18, 17, 16, and 10
clusters in D1--4. This optical cluster catalog was chosen over those of other selection
techniques as it provides consistent completeness for all four SNLS Deep field areas. 

Olsen et al. (2007) find their cluster redshifts are slightly overestimated for $z < 0.6$, and
underestimated for $z > 0.7$, with a standard deviation of $\sim0.1$
\citep{LFO}. For the cluster redshift we instead use the peak of the redshift
distribution from Ilbert catalog E/S0 and Sbc galaxies along the cluster's line of sight
(within 40 arcseconds).  For distributions which plateau instead of peak,
the plateau's central redshift is used. To evaluate corrected cluster redshift precision we apply
the same procedure to the VVDS spectroscopic redshift catalog\footnote{http://www.oamp.fr/virmos/}.
For the four D1 clusters with VVDS galaxies along their line of sight, the
differences between photometric and spectroscopic redshift peaks have a
standard deviation of $\sigma_{\Delta z / (1+z_{VVDS})}=0.023$, indicating
good agreement.

\subsection {Additional Catalog Processing}
\label{sS2add} 

SNe\,Ia host galaxies are identified as the closest neighbor in the Ilbert catalog - except
when the offset difference between the nearest two is less than 2 arcseconds, in which case
redshift is used to discriminate. 85 SNe\,Ia have no neighbor within 5 arcseconds (maximum host offset),
and cannot be used; %and must be considered hostless;
this mainly includes SNe\,Ia in hosts of $i_M>25$, and on the masked regions covering $\sim$20\% of field area.
For the SNe\,Ia with potential Ilbert catalog hosts, iterative outlier rejection 
is applied to the residual dispersion between host photometric and SN\,Ia spectroscopic redshifts\footnote{A sigma of three and fractional convergence threshold of 0.01 was used; average number of iterations was 4}.
The photometric redshift accuracy when performed for all fields combined is $\sigma_{\Delta z/(1+z_{SN})} = 0.030$ (consistent with Ilbert et al. 2007).
However, if outlier rejection is applied to each Deep field separately, we find that
each has a different photometric redshift uncertainty: $\sigma_{D1}=0.028, \sigma_{D2}=0.024, \sigma_{D3}=0.030, \sigma_{D4}=0.050$.
In this way, 26 SNe\,Ia are rejected as outliers, dropping the total number to 232 SNe\,Ia.

The SN\,Ia rate per year, $SNR_{Ia}$, is calculated for every galaxy
based on the two component model, parametrized
by $SNR_{Ia} = A\,M + B\,\dot{M}$ where $M$ is the stellar
mass of the galaxy (in solar masses) and $\dot{M}$ is the star formation
rate (in solar masses per year). The most recent A and B values of
Sullivan et al. (2006a) are used: $\rm A = 5.3 \pm 1.1 \times 10^{-14} \ SNe \ y^{-1} \ M_{\odot}^{-1}$ and
$\rm B = 3.9 \pm 0.7 \times 10^{-4} \ SNe \ y^{-1} \ (M_{\odot} \ y^{-1})^{-1}$.
The expected SN\,Ia rate per year for each galaxy is corrected to
reproduce the total number observed in the SN\,Ia
sample (excluding SNe\,Ia with no identified host). This correction is performed for
separate redshift ranges and Deep fields to account for incompleteness in
observations; values are given in Table \ref{TS2norms}. The result is the number of SN\,Ia expected in
each galaxy over the observing period. These calculations are also
repeated using the $A$ (mass) component only.

%%%%%%%%%%%%%%%%%%%%%%%%%%%%%%%%%%%%%%%%%%%%%%%%%%%%%%%%%%%%%%%%%%%%%%%%%%%%%%%%%%%%%%%%%%%%%%%%%%%%%%%%%%%%%%%%%%%%%%%
 
\section{PARAMETRIZED CLUSTER-LIKE ENVIRONMENTS}
\label{S3SigParam}

To begin we use a continuous clustering strength estimator to compare the
SN\,Ia rate per unit mass in and out of clustered environments.
This parameter quantifies the significance ($\Sigma$) of finding $N_E$ neighbor
galaxies the environment of a galaxy or SN\,Ia, given number expected from the background distribution.
$\Sigma$ is calculated for a cylindrical volume environment of diameter $D$ and redshift
depth $\pm \sigma_{\Delta z/(1+z)}(1+z)$ ($\sigma$ from \S~\ref{sS2add})
centered on the object, as in equation \ref{ES3sig} where 
$N_E$ is the number of galaxies in the environment, 
$N_F$ is the number of galaxies within $\pm \sigma_{\Delta z/(1+z)}(1+z)$ in the Deep field,
$A_E = \pi D^2/4$ is the aperture area,
and $A_F$ is the area of the field after allowing for the area
occulted by foreground star masks \citep{I06}.
Since $N_F$ is computed at the same redshift
as the object being studied, incompleteness at high redshift is 
automatically compensated for. 

\begin{equation}
  \Sigma = \frac{N_E-N_F(A_E/A_F)}{\sqrt{N_F(A_E/A_F)}}
\label{ES3sig}
\end{equation}
 
\begin{figure}
\begin{center}
  \includegraphics[scale=.80]{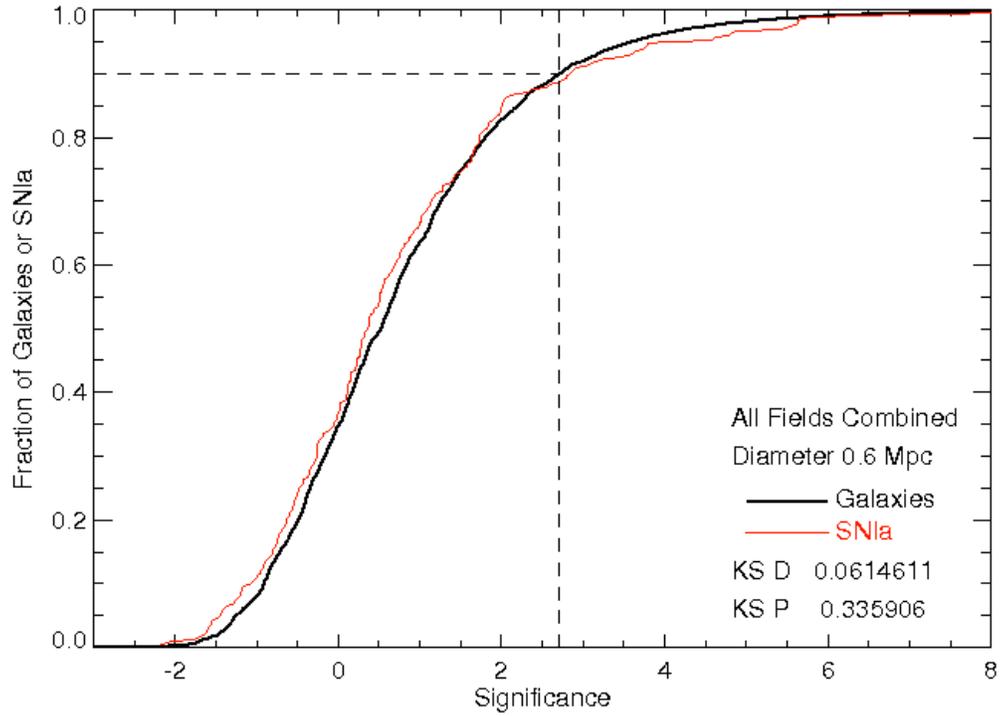}
%  \plotone{f1.ps}
  \caption{ Cumulative significance distribution for galaxies and SNe\,Ia, with environment diameter 0.6 Mpc, for D1--4. Kolmogorov-Smirnov maximum difference (KS D) and probability (KS P) show the null hypothesis cannot be ruled out. Dashed lines represent $\Sigma_{10\%}$. \label{FS3sigdistr}}
\end{center}
\end{figure}

The advantage of this parametrization is that a strict cluster definition is avoided - any
desired scale of clustering can be explored by altering the aperture
diameter $D$.
Environmental significances are calculated using a range of clustering
scales (diameters $D$ ranging from 0.1 to 1.5 Mpc) both for field
galaxies ($0.1 < z < 1.1$), and also for the 232 SNe\,Ia surviving the
cuts in section \ref{sS2add}. Normalized cumulative significance distributions
for both field galaxies and SNe\,Ia are shown in Figure
\ref{FS3sigdistr} for $D=0.6$ Mpc. A Kolmogorov-Smirnov test shows
that the two samples are statistically indistinguishable: SN hosts
appear to be drawn from the same population as field galaxies with respect
to clustering in their environment.

Let us now turn to
the most clustered environments, by defining a significance
limit ($\Sigma_{X\%}$) to isolate the top 10\%, 5\%, and 1\% most
significant galaxy environments -- the high significance group
(HSG). Figure \ref{FS3sigdistr} shows the HSG cutoffs for
$\Sigma_{10\%}$ as an example; actually the HSG is determined
for each diameter $D$ and each Deep field separately, then the HSGs for
all are fields combined for a given $D$.
We use summed Poisson probabilities to compare the number of SNe\,Ia
in the HSG ($N_{obs}$) to the number expected ($N_{exp}$), which is the sum
of the expected number of SNe\,Ia in each HSG galaxy (from \S~\ref{sS2add}).
The Poisson probability of observing $x=N_{obs}$ given $\mu=N_{exp}$ is expressed
by equation \ref{ES3PP} \citep{Bev}, and the summed Poisson probabilities 
for $x >\mu$ and $x < \mu$ are given in
equations \ref{ES3PPS} and \ref{ES3PPS2} respectively.

\begin{equation}
  \label{ES3PP}
  P_P(x;\mu) = \mu^{x} \frac{e^{-\mu}}{x!}
\end{equation}
 
\begin{equation}
  \label{ES3PPS}
  P_{SUM}(x,\infty; \mu) =  \sum_{x'=x,x+1,...}^{x'=\infty} P_P(x';\mu)
\end{equation}
 
\begin{equation}
  \label{ES3PPS2}
  P_{SUM}(0,x; \mu) =  \sum_{x'=0,1,...}^{x'=x-1} P_P(x';\mu)
\end{equation}

As expressed by the morphology-density relation, galaxy clusters are dominated
by early-type galaxies \citep{Postman84}, so Poisson probabilities are also
calculated for the subset of early-type field galaxies and SNe\,Ia hosts in the HSG.
Results for a representative sample of environment diameters in Table
\ref{TS3probs} show the number of HSG SN\,Ia to be consistent with
expectations of the two-component model over a range of $D$ and
$\Sigma_{X\%}$. Thus we conclude this clustering parametrization
method does {\it not} show an influence of clustering on SN\,Ia events.
Neither including the SNe\,Ia outliers rejected in \S~\ref{sS2add},
nor increasing the environment redshift depth to $\pm 2 \sigma (1+z)$,
affects this conclusion.

%%%%%%%%%%%%%%%%%%%%%%%%%%%%%%%%%%%%%%%%%%%%%%%%%%%%%%%%%%%%%%%%%%%%%%%%%%%%%%%%%%%%%%%%%%%%%%%%%%%%%%%%%%%%%%%%%%%%%%%

\section{SN\,Ia IN OLSEN CATALOG CLUSTERS}
\label{S4LFO}
 
Here we identify SNe\,Ia and galaxies in grade A clusters from the Olsen et al. (2007)
cluster catalog. We use Poisson probabilities (\S~\ref{sS4probs}) and a
direct calculation of SNe\,Ia rates in clusters (\S~\ref{sS4rates}) to look for an
influence of clustering on the SN\,Ia rate per unit mass.
As the SNLS detection efficiencies of Neill et al. (2006) are valid for $0.2 < z < 0.6$,
we restrict cluster, galaxy, and SN\,Ia redshifts to this range to use of them.
This decreases the catalogs to 30 clusters, 70587 galaxies, and 109 SNe\,Ia.
We note Olsen et al. (2007) and Ilbert et al. (2006) use different Terapix data releases
(T0002 and T0003 respectively), and the release used by Olsen et al. has more masked regions.
However the differences in field effective areas are $\lesssim 10\%$, so likely only
$\lesssim 3$ clusters of Ilbert catalog galaxies are missing from the Olsen catalog.

\begin{figure}
\begin{center}
%  \epsscale{0.8}
%  \plottwo{f2a.eps}{f2b.eps}
%  \epsscale{0.8}
%  \plottwo{f2c.eps}{f2d.eps}
%  \epsscale{0.35}
%  \plotone{f2e.eps}
  \includegraphics[scale=.30]{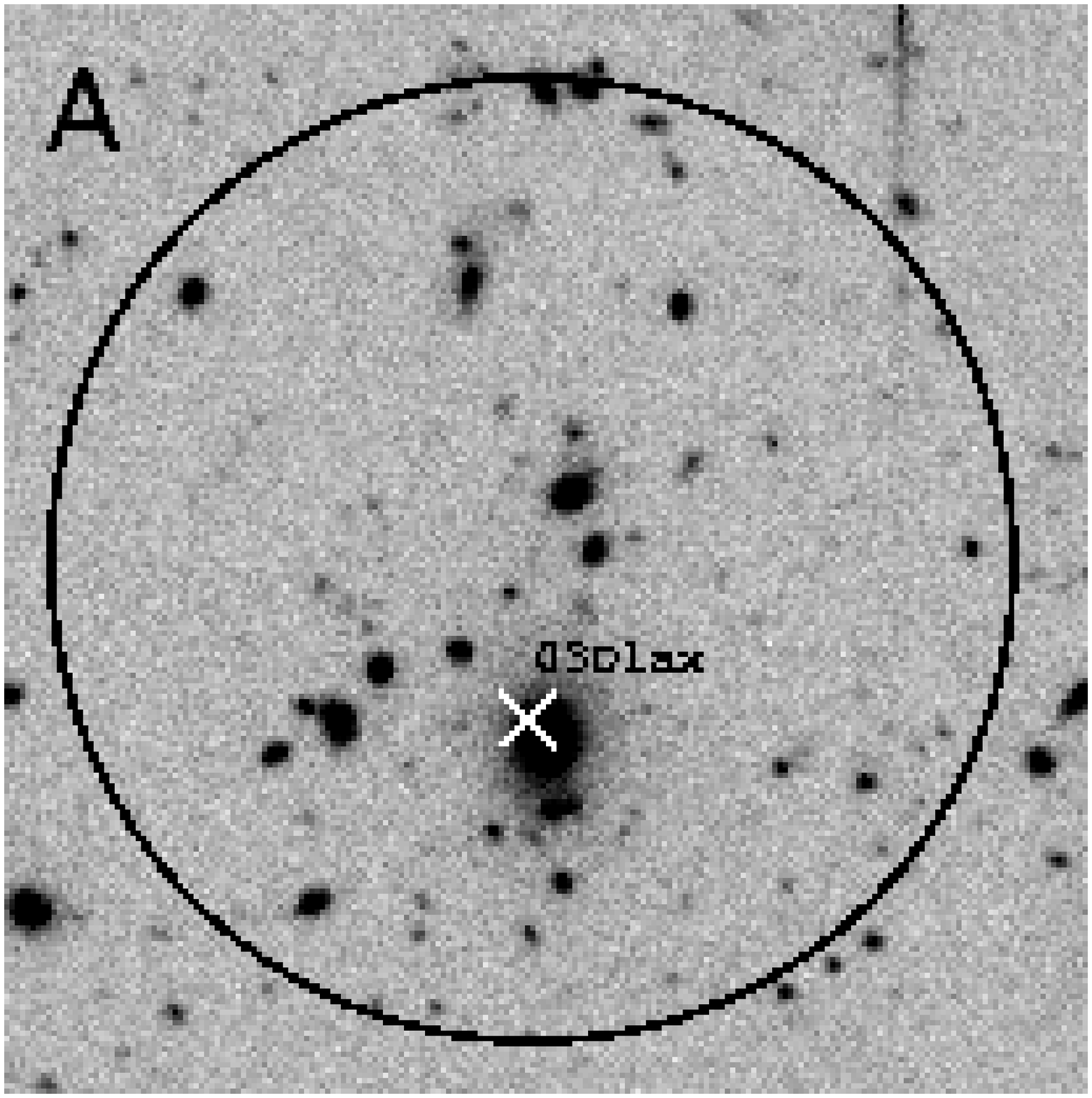}
  \includegraphics[scale=.30]{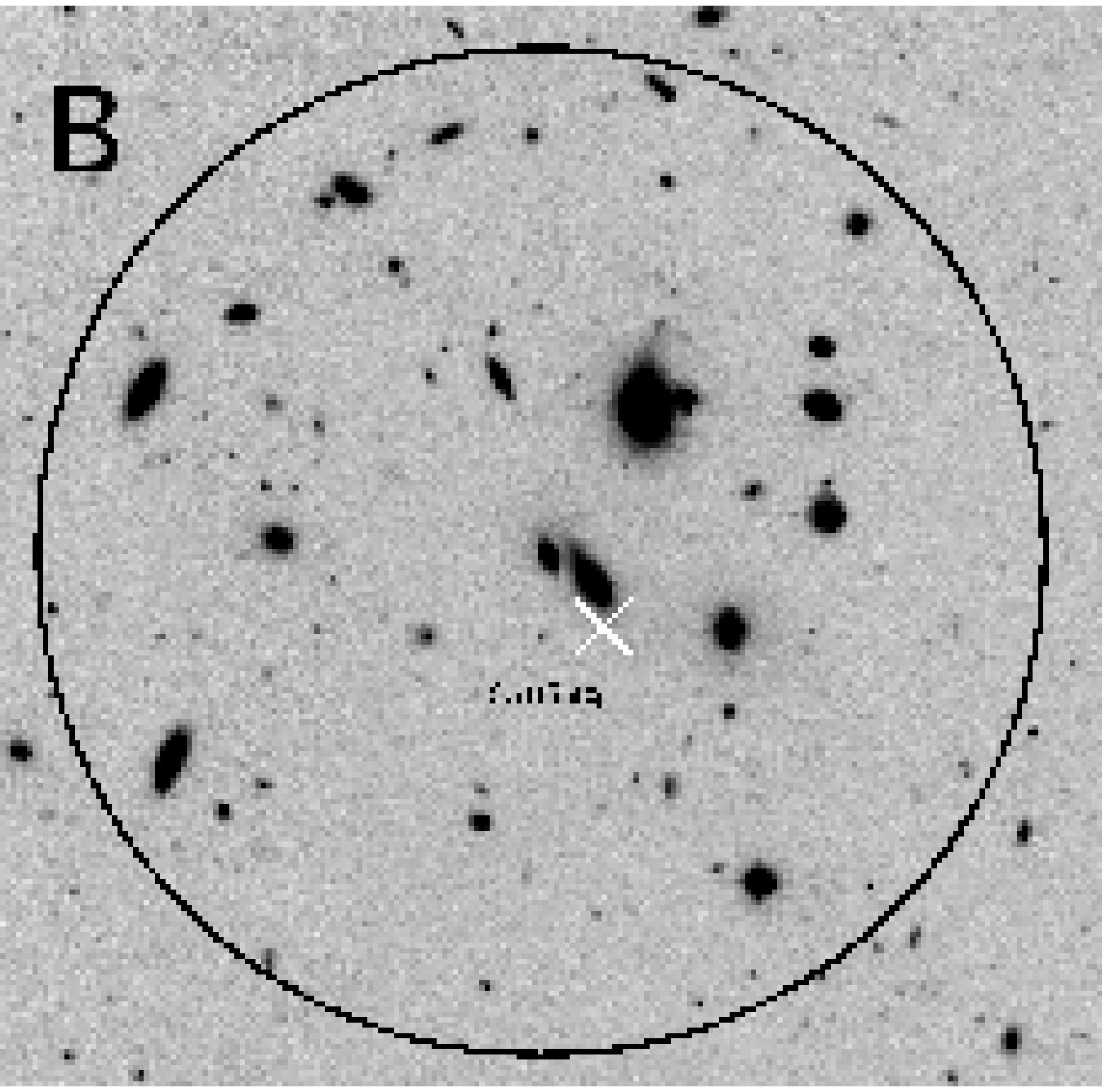}
  \includegraphics[scale=.35]{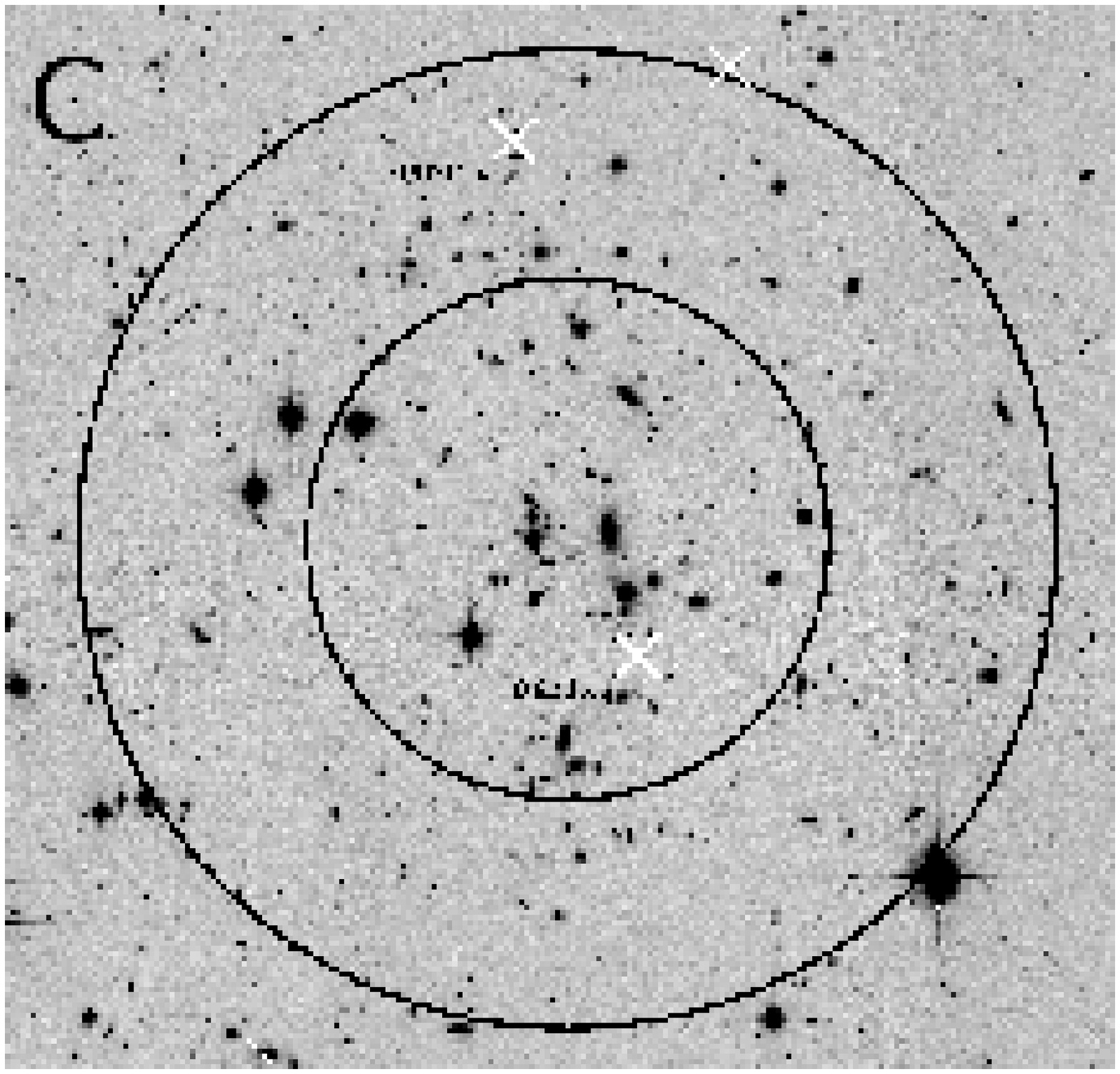}
  \includegraphics[scale=.35]{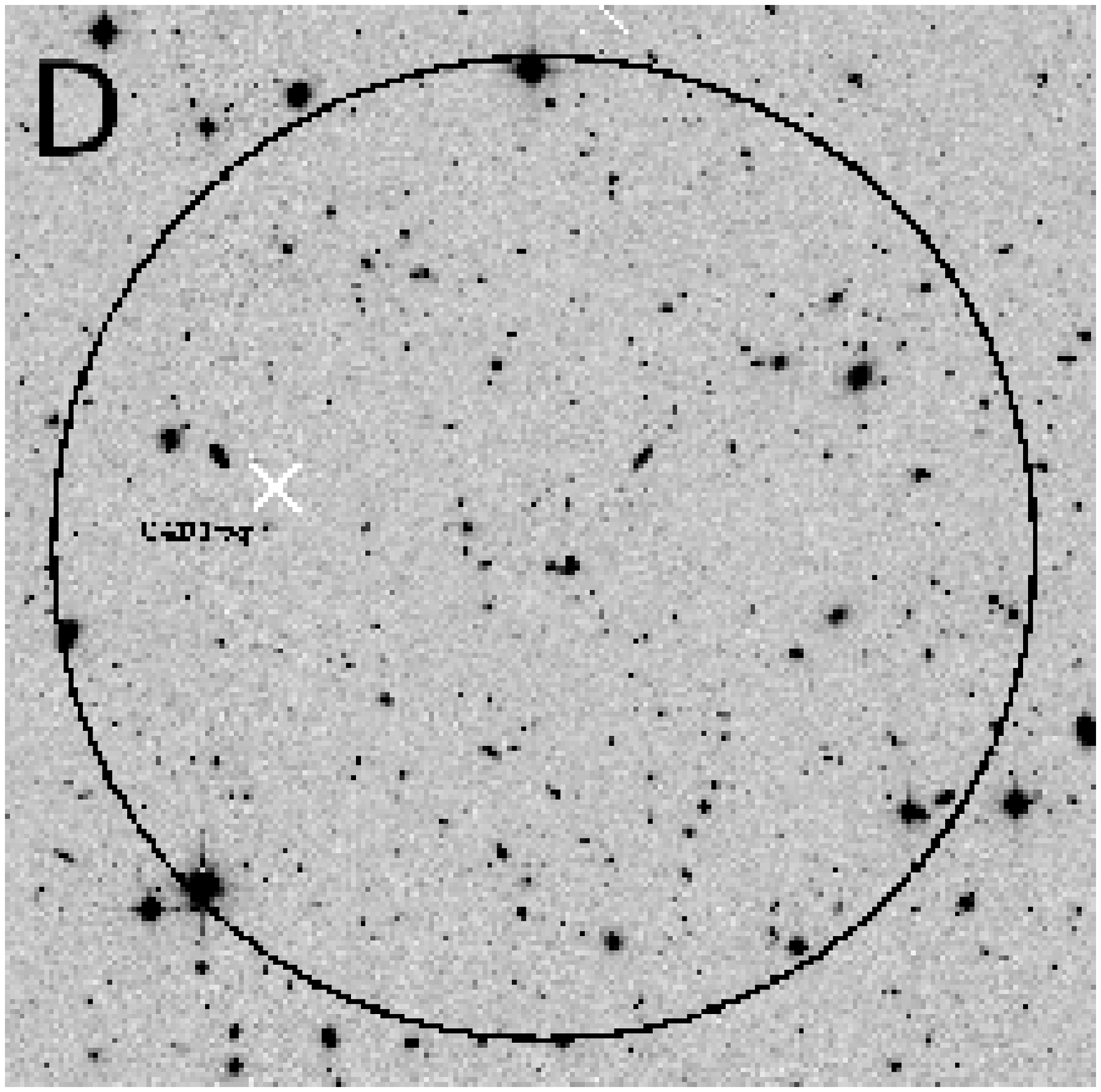}
  \includegraphics[scale=.35]{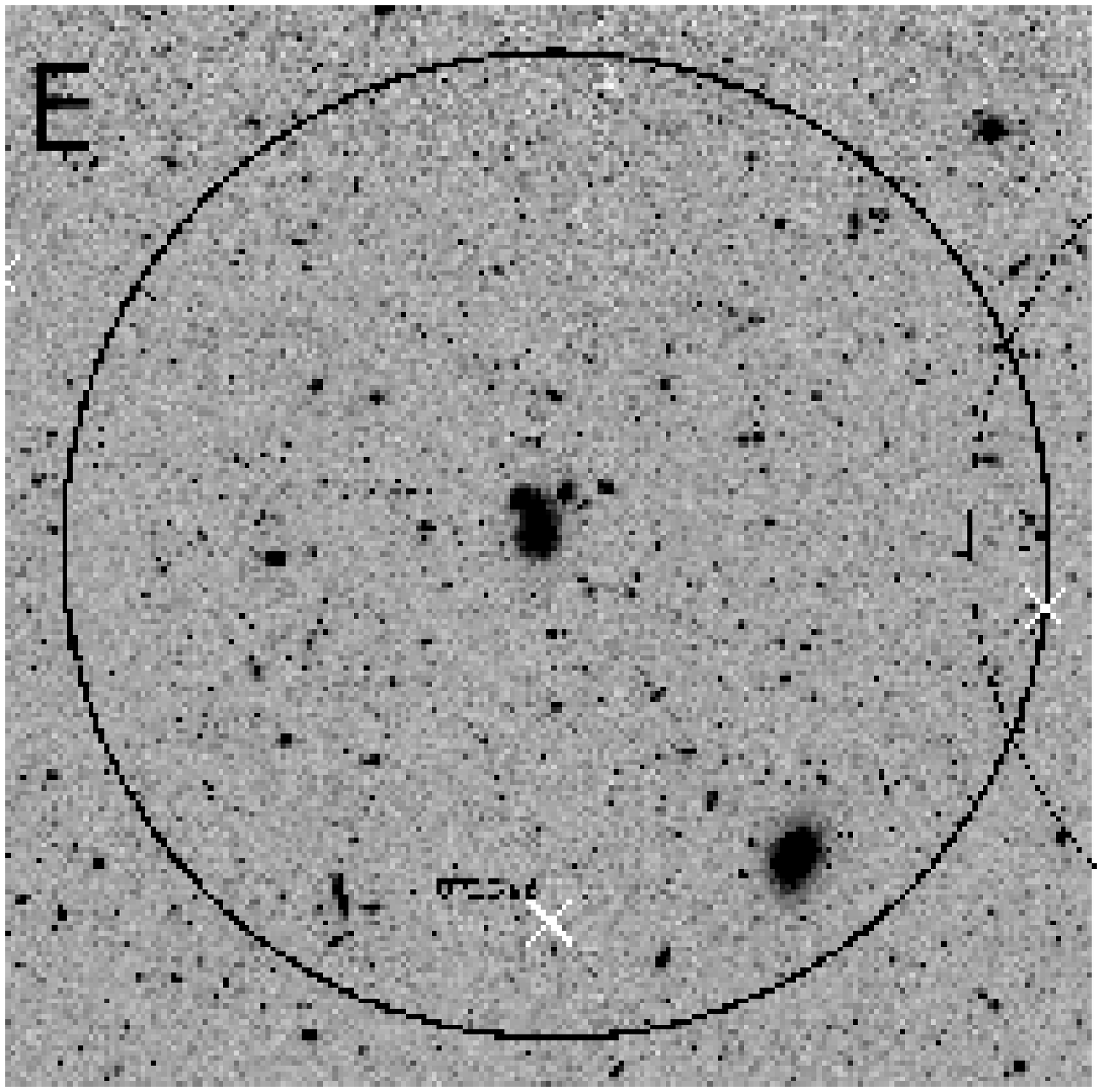}
  \caption{ SNLS SNe\,Ia identified in galaxy clusters from the Olsen et al. (2007) catalog.
    SNe\,Ia coordinates marked by crosses, clusters represented by circles of diameter 0.4 Mpc (A, B), 0.8 Mpc (inner circle of C),
    and 1.5 Mpc (outer circle of C, D, and E). Images created from SNLS 2004 reference images and Skycat (image quality degraded for submission to astro-ph).
    \label{FS4images}}
\end{center}
\end{figure}

SNe\,Ia and galaxy members of clusters are identified as are
neighbors in the environment of a galaxy in \S~\ref{S3SigParam},
except the volume is centered on the cluster coordinates.
Since the cluster filter used by Olsen et al. (2007) has a profile with core radius
$r_c=0.133h^{-1}_{75}$ Mpc and cut off radius of $r_{co}=1.33h^{-1}_{75}$ Mpc,
results for 0.4, 0.8, and 1.5 Mpc will be presented as representative.
For galaxies a redshift depth of 
$\pm 2 \sqrt{ (\sigma_{\Delta z/(1+z_{SN})})^2 + (\sigma_{\Delta z / (1+z_{VVDS})})^2 }\,(1+z_C)$
is used, a convolution of uncertainties in cluster redshifts from \S~\ref{sS2LFO} and galaxy 
redshifts from \S~\ref{sS2add}. As SNe\,Ia have spectroscopic redshifts, a redshift depth of
simply $\pm2\sigma_{\Delta z / (1+z_{VVDS})}$ is appropriate.
Images in Figure \ref{FS4images} and data in Table \ref{TS4clustSN}
present 6 SNe\,Ia identified in Olsen clusters. While the 3 SNe\,Ia within $D=0.8$ Mpc
are probably physically associated with the clusters, this cannot be said
for the remaining three. The number of field SNe\,Ia with $0.2<z_{SN}<0.6$ predict
$\sim1.8$ SNe\,Ia would randomly appear in the regions between $r=0.4$ and $r=0.75$ Mpc
and $\pm2\sigma_{\Delta z / (1+z_{VVDS})}$. The probabilities of all 3 being random
and all 3 {\it not} being random associates are both $<20\%$. As it remains likely that at least one is physically
associated with a cluster we include them in our results, but remind the reader that interloping SNe\,Ia
(and galaxies) will add to the uncertainties for $D=1.5$ Mpc.

\subsection{Summed Poisson Probabilities}
\label{sS4probs}
 
Summed Poisson probabilities (defined in \S~\ref{S3SigParam})
of observing these cluster SNe\,Ia are computed from the number
expected in the identified member galaxies, for all galaxy types
and early-types only.
Table \ref{TS4probs} shows these observations are consistent with
the two-component SN\,Ia rate model.  In fact, $\geq6$
SNe\,Ia would have to have been observed within $D=$ 0.4 Mpc of galaxy
clusters ($\geq8$ or 0 for $D=$ 0.8 Mpc) for $P_{SUM} < 0.05$.
This constrains the SN\,Ia rate in clusters to agree with the
two-component model to within a factor of two.

Obviously, to detect an effect on SNe\,Ia rates due to clustering (over
and above the expectations of the two component model and morphology-density
relation) lurking in this data, we should at least detect the
two component model. So can we rule out the single component model
(mass or ``A'' component only) from these cluster
observations? The answer appears to be no.
Table \ref{TS4probs} shows that the number of observed cluster SN\,Ia is
actually consistent with {\it both} models. Surveys to test for the A+B model
or more complicated delay time distributions \citep{Mann06} will require a
larger survey (two to three times the area or duration); unfortunately,
the final SNLS data set will not be adequate for this.

\subsection{SN\,Ia Rates in Clusters}
\label{sS4rates}
 
Here we use detection efficiencies from Monte Carlo simulations of the
SNLS SN\,Ia identification pipeline 
from Neill et al. (2006) to directly calculate the SN\,Ia rate per unit
mass in clusters, and compare it to that for the field.
The corrected number of SN\,Ia which exploded in a given field in a year,
$N_{corr,Ia}$, is extracted from equation 3 of Neill et al. (2006).
As shown in equation \ref{ES4NcorrIa}, 
$N_{Ia}$ is the total number of supernova observed in a given field;
$S$ is the number of observing seasons;
$\epsilon_{yr}$ is the detection efficiency per year (the probability of a supernova being
detected, sent for spectroscopy, and identified as a type Ia);
and $C_{SPEC}$ accounts for the fraction of SNe\,Ia for which spectra are
obtained yet remain unidentified.
$[1+\langle z\rangle_V]$ corrects for time dilation at the volume-weighted
average redshift of the survey, and $\langle z\rangle_V =$ 0.46.
Detection efficiencies from Neill et al. (2006) and the number of observing seasons
for each field up to January 2007 are in Table \ref{TS4Ncorrparams}.
Approximations to Poisson uncertainty from Gehrels (1986) determine the upper and
lower limits on $N_{Ia}$ at the 0.84 confidence level (corresponding to $1\sigma$),
which are substituted into equation \ref{ES4NcorrIa} to determine $\Delta N_{corr,Ia}$.

\begin{equation}
\label{ES4NcorrIa}
N_{corr,Ia} = \frac{N_{Ia}/S}{\epsilon_{yr} C_{SPEC}} [1+\langle z\rangle_V]
\end{equation}
 
Table \ref{TS4NcorrIa} contains the resulting corrected SN\,Ia rate
per unit mass per year in clusters and the field. With only 3 identified
cluster SNe\,Ia, statistical uncertainties dominate this calculation; systematics, mainly
the error in galaxy mass calculations, are likely another $\sim$30\%.
Thus, our SN\,Ia rate is consistent with both the 
rate in early-type galaxies, $\rm 5.3\pm 1.1 \times 10^{-14} \ SNe
\ y^{-1} \ M_{\odot}^{-1}$ \citep{S06a}, and the low redshift cluster rate
from WOOTS, $\rm 9.8^{+5.9}_{-3.9}\pm 0.9
\times 10^{-14} \ SNe \ y^{-1} \ M_{\odot}^{-1}$ \citep{KS07}.

There are two factors not considered which could affect SN\,Ia
detection efficiencies in cluster galaxies.  First,
SN\,Ia detection efficiencies decrease in brighter hosts \citep{N06},
and the brightest galaxies are early-type.  Second, the SN\,Ia
detection efficiency decreases for fainter, lower stretch SN\,Ia, and
these faint SN\,Ia occur preferentially in early-type hosts
\citep{S06a}. Since cluster galaxies are predominantly early-type, both
of these effects dominate in clusters: the first effect decreases
the number expected in clusters by $\sim 15\%$, but quantifying the second
would require more detailed completeness simulations.
Both effects would cause us to underestimate the
rate of SNe\,Ia in clusters relative to the field.

\subsection{SN\,Ia Rates in E/S0 Cluster Galaxies}
\label{sS4ES0rates}

To avoid these effects and the morphology-density relation, 
we limit the galaxy sample to two subsets: all early-type galaxies, and the brightest population of early-type galaxies
(those with $M_V<-23.0$ like brightest cluster galaxies, BCGs).
This has the added benefit of rejecting interlopers misidentified as cluster members.
Two cluster SNe\,Ia have early-type hosts, and the host of 03D1ax is brighter than
$M_V=-23.0$. Detection efficiency corrections are performed as described above, 
with the final results and rates presented in Table \ref{TS4NcorrIa}.

Although these samples are more sensitive to the two detection
efficiency biases affecting early-type galaxies (\S~\ref{sS4rates}), limiting all galaxies
to early-types minimizes differences between the clusters and the field and results in a more
meaningful test. The results, although not statistically significant, are {\it suggestive}
of the rate enhancement in cluster over field early-type galaxies established by
Mannucci et al. (2007). An enhancement in BCG-like galaxies would be consistent
with the findings of an enhanced SN\,Ia rate in radio loud elliptical
galaxies \citep{DV05}, as these are usually the brightest cluster
galaxies. Future work will investigate the rates of SNLS SNe\,Ia in
radio galaxies using existing radio catalogs for the Deep fields.

\subsection{Effects of Altering Data Constraints}
\label{sS4robust}

Including grade B clusters from the Olsen et al. (2006) catalog increases the total number
of clusters to 65, with one new cluster SN\,Ia identified (03D1fb at $z_{SN}=0.498$ in an E/S0 host of $M_V=-21.9$ with cluster offset $57.8''$).
This does not affect the conclusions of the Poisson probability experiment (\S~\ref{sS4probs}).
The resulting cluster rates for $D=0.8$ Mpc for all galaxy types, early-types, and the brightest early-types only are
$6.7^{+8.5}_{-3.0}$,
$6.7^{+11.0}_{-3.4}$, and
$6.2^{+29.2}_{-5.1}$ $\rm \times 10^{-14} \ SNe \ y^{-1} \ M_{\odot}^{-1}$;
slightly below yet well within uncertainties of results with A grade clusters only.

None of the SNe\,Ia rejected as outliers in \S~\ref{sS2add} are associated with Olsen catalog clusters.
Including them does not affect the conclusions of the Poisson probability experiment,
and only increases the SNe\,Ia field rate for all galaxy types to $15.9^{+1.8}_{-1.5}$
$\rm \times 10^{-14} \ SNe \ y^{-1} \ M_{\odot}^{-1}$. Including in the experiment all SNe\,Ia
for which no host was identified does not yield any new cluster SNe\,Ia either.

Reducing the SNe\,Ia sample to those included in Neill et al. (2007) avoids a
possible over-correction for detection efficiencies which might arise from extending the sample
to later times, if the survey became more efficient (we estimate any improvement in
detection efficiency to be small). This decreases the number of SNe\,Ia
to 40, with one cluster SNe\,Ia (03D1ax). This does not alter the overall conclusions of these experiments. 

Lastly, the inclusion of SNLS SNe\,Ia candidates which did not receive spectroscopic confirmation
(past maximum light, or detected as field season was ending), but which were given
photometric types and redshifts using the techniques of Sullivan et al. (2006b),
does not yield any new cluster SNe\,Ia or affect the conclusions of these experiments.

%%%%%%%%%%%%%%%%%%%%%%%%%%%%%%%%%%%%%%%%%%%%%%%%%%%%%%%%%%%%%%%%%%%%%%%%%%%%%%%%%%%%%%%%%%%%%%%%%%%%%%%%%%%%%%%%%%%%%%%
 
\section{CONCLUSIONS}
\label{S5Conc}

Type Ia supernova, galaxy, and cluster catalogs generated from the
first four years of CFHTLS Deep survey data were combined to search for
an influence of clustering on the SN\,Ia rate per unit mass.
To avoid cluster-specific detection efficiencies and the
inclusion of interloping galaxies as cluster members, we 
also considered subsets of regular and BCG-like early-type
galaxies only. Results are dominated by the statistical uncertainties
in identifying only three probable and three possible cluster SNe\,Ia, and consistent
with the results of low redshift cluster SNe\,Ia rate studies.
To capitalize on SNLS's inclusion of both field and cluster environments,
we used the continuous clustering strength parameter ``significance'' to compare the
SN\,Ia rate per unit mass in and out of clustered environments, but did
not find evidence of a clustering influence on the SN\,Ia rate per unit mass.
Future work includes incorporating existing radio and infrared catalogs to investigate
the rates of SNLS SNe\,Ia in radio loud and infrared luminous galaxies,
and the VIRMOS spectroscopic redshifts for Deep field galaxies to
explore the SNe\,Ia rate in small groups.

%%%%%%%%%%%%%%%%%%%%%%%%%%%%%%%%%%%%%%%%%%%%%%%%%%%%%%%%%%%%%%%%%%%%%%%%%%%%%%%%%%%%%%%%%%%%%%%%%%%%%%%%%%%%%%%%%%%%%%%
 
\acknowledgments
 
We gratefully acknowledge the CFHT Queued Service Observations team,
Olivier Ilbert and Henry McCracken for early access to and correspondence regarding their photometric redshift galaxy catalog,
Lisbeth Olsen for early access the optical cluster catalog,
Jon Willis and Dan Maoz for helpful conversations,
and our anonymous referee for constructive correspondence.
This paper is based on observations obtained with MegaPrime/MegaCam, a joint project of CFHT and CEA/DAPNIA, at the CFHT, which is operated by the National Research Council of Canada, the Institut National des Science de l'Univers of the Centre National de la Recherche Scientifique (CNRS) of France, and the University of Hawaii.
This paper is also based on spectroscopic observations obtained at the Gemini Observatory which is operated by the Association of Universities for Research in Astronomy, Inc., under a cooperative agreement with the NSF on behalf of the Gemini partnership;
the Very Large Telescope at the European Southern Observatory in Paranal, Chile;
the W. M. Keck Observatory which is operated as a scientific partnership among the California Institute of Technology, the University of California, and the National Aeronautics and Space Administration;
and the 6.5 meter Magellan Telescopes located at Las Campanas Observatory, Chile.
This work is based in part on data products from the Canadian Astronomy
Data Centre as part of the CFHT Legacy Survey, a collaborative project
of NRC and CNRS.
This work has been supported by NSERC and the University of Victoria.

%%%%%%%%%%%%%%%%%%%%%%%%%%%%%%%%%%%%%%%%%%%%%%%%%%%%%%%%%%%%%%%%%%%%%%%%%%%%%%%%%%%%%%%%%%%%%%%%%%%%%%%%%%%%%%%%%%%%%%%
 
{\it Facilities:} \facility{CFHT}.

%%%%%%%%%%%%%%%%%%%%%%%%%%%%%%%%%%%%%%%%%%%%%%%%%%%%%%%%%%%%%%%%%%%%%%%%%%%%%%%%%%%%%%%%%%%%%%%%%%%%%%%%%%%%%%%%%%%%%%%

%\clearpage 

%%%%%%%%%%%%%%%%%%%%%%%%%%%%%%%%%%%%%%%%%%%%%%%%%%%%%%%%%%%%%%%%%%%%%%%%%%%%%%%%%%%%%%%%%%%%%%%%%%%%%%%%%%%%%%%%%%%%%%%

%%%%%%%%%%%%%%%%%%%%%%%%%%%%%%%%%%%%%%%%%%%%%%%%%%%%%%%%%%%%%%%%%%%%%%%%%%%%%%%%%%%%%%%%%%%%%%%%%%%%%%%%%%%%%%%%%%%%%%%
 
% Tables
 
\clearpage
\begin{table}
  \begin{center}
    \caption{SN\,Ia rate correction factors. \label{TS2norms}}
    \begin{tabular}{ccccc}
      \tableline\tableline
      Redshift Range & D1 & D2 & D3 & D4 \\
      \tableline
      0.0-0.5  & 1.431  & 1.016  & 1.538  & 0.896 \\
      0.5-0.7  & 1.203  & 0.511  & 0.940  & 0.999 \\
      0.7-0.9  & 0.523  & 0.194  & 0.418  & 0.616 \\
      0.9-1.1  & 0.078  & 0.065  & 0.140  & 0.118 \\
      1.1-1.3  & 0.000  & 0.000  & 0.000  & 0.000 \\
      \tableline
    \end{tabular}
  \end{center}
\end{table}

\clearpage
\begin{table}
  \begin{center}
    \caption{Results of the clustering parametrization method for D1--4. \label{TS3probs}}
    \begin{tabular}{cccccccccc}
      \tableline\tableline
      Diameter    &  \multicolumn{3}{c}{$\Sigma_{10\%}$}  &  \multicolumn{3}{c}{$\Sigma_{5\%}$} & \multicolumn{3}{c}{$\Sigma_{1\%}$} \\
      (Mpc)       &  $N_{obs}$ & $N_{exp}$ & $P_{SUM}$ & $N_{obs}$ & $N_{exp}$ & $P_{SUM}$ & $N_{obs}$ & $N_{exp}$ & $P_{SUM}$ \\
      \tableline
      0.4          &  21 & 26.9  & 0.15   &  11  & 15.0   & 0.18   & 3  & 3.98   & 0.44 \\
      0.6          &  27 & 26.5  & 0.48   &  16  & 14.7   & 0.41   & 5  & 3.41   & 0.26 \\
      0.8          &  21 & 25.3  & 0.23   &  12  & 13.8   & 0.38   & 5  & 3.15   & 0.21 \\
      1.0          &  23 & 24.8  & 0.41   &  10  & 13.2   & 0.23   & 6  & 3.13   & 0.10 \\
      1.5          &  21 & 22.3  & 0.44   &  11  & 12.0   & 0.46   & 2  & 2.98   & 0.43 \\
      \tableline
      \multicolumn{10}{l}{For early-type galaxies and SN\,Ia hosts only:} \\
      0.4 & 5 & 7.43 & 0.25 & 3 & 3.77 & 0.48 & 0 & 0.72 & 0.49 \\
      0.6 & 6 & 6.86 & 0.47 & 3 & 3.27 & 0.59 & 0 & 0.60 & 0.55 \\
      0.8 & 5 & 6.33 & 0.39 & 3 & 3.08 & 0.63 & 1 & 0.49 & 0.39 \\
      1.0 & 3 & 5.98 & 0.15 & 3 & 2.86 & 0.54 & 1 & 0.59 & 0.45 \\
      1.5 & 2 & 5.72 & 0.08 & 2 & 2.76 & 0.48 & 0 & 0.61 & 0.54 \\
      \tableline
    \end{tabular}
  \end{center}
\end{table}

\clearpage
\begin{table}
  \begin{center}
    \caption{Cluster SNe\,Ia details. \label{TS4clustSN}}
    \begin{tabular}{cccccccccc}
      \tableline\tableline
      SN\,Ia  & SN\,Ia        & SN\,Ia        & SN\,Ia     & SN\,Ia     & host & host   & host   & cluster    & cluster  \\
      SNLS ID & RA            & Dec.          & $z_{spec}$ & stretch    & type & $M_V$  & offset & $z_{phot}$ & offset   \\
      \tableline
      03D1ax  & $\rm 02^h 24^m 23^s.32$  & $\rm -04^{\circ} 43' 14''.41$  & 0.496 & $\ldots$ & ES/0 & -23.64 & $1.97''$ & 0.53 & $10.8''$   \\
      06D1kg  & $\rm 02^h 24^m 32^s.57$  & $\rm -04^{\circ} 15' 02''.0$   & 0.323 & 1.21     & Sbc  & -20.31 & $1.69''$ & 0.30 & $47.2''$   \\
      05D1by  & $\rm 02^h 24^m 35^s.45$  & $\rm -04^{\circ} 12' 04''.2$   & 0.299 & 0.99     & Sbc  & -21.17 & $0.86''$ & 0.30 & $138.6''$  \\
      04D1pg  & $\rm 02^h 27^m 04^s.16$  & $\rm -04^{\circ} 10' 31''.4$   & 0.515 & 1.05     & Sbc  & -19.62 & $0.18''$ & 0.51 & $85.2''$   \\
      05D3mq  & $\rm 14^h 19^m 00^s.40$  & $\rm +52^{\circ} 23' 06''.81$  & 0.246 & 0.90     & ES/0 & -21.85 & $4.94''$ & 0.25 & $9.99''$   \\
      07D3af  & $\rm 14^h 19^m 05^s.01$  & $\rm +53^{\circ} 06' 08''.98$  & 0.356 & 0.98     & Scd  & -18.61 & $0.35''$ & 0.30 & $128.7''$  \\
      \tableline
    \end{tabular}
  \end{center}
\end{table}

\clearpage
\begin{table}
  \begin{center}
    \caption{Summed Poisson probabilities for cluster SNe\,Ia for D1--4. \label{TS4probs}}
    \begin{tabular}{cccccc}
      \tableline\tableline
      Cluster   &              & \multicolumn{2}{c}{A+B Components} & \multicolumn{2}{c}{A Component} \\
      Diameter  &   $N_{obs}$  &  $N_{exp}$  & $P_{SUM}$   &  $N_{exp}$  & $P_{SUM}$   \\
      \tableline
      0.4 & 2 & 2.38 & 0.57 & 3.21 & 0.38  \\
      0.8 & 3 & 3.51 & 0.54 & 4.53 & 0.34  \\
      1.5 & 6 & 5.97 & 0.55 & 7.30 & 0.41  \\
      \tableline
      \multicolumn{6}{l}{For early-type galaxies only:}    \\
      0.4 & 2 & 1.97 & 0.59 & 2.79 & 0.47  \\
      0.8 & 2 & 2.60 & 0.52 & 3.67 & 0.29  \\
      1.5 & 2 & 3.69 & 0.29 & 5.23 & 0.11  \\
      \tableline
    \end{tabular}
  \end{center}
\end{table}

\clearpage
\begin{table}
  \begin{center}
    \caption{Parameter values for equation \ref{ES4NcorrIa}. \label{TS4Ncorrparams} }
    \begin{tabular}{cccc}
      \tableline\tableline 
      Field & Seasons & $\epsilon_{yr}$\tablenotemark{a} & $C_{SPEC}$\tablenotemark{a} \\
      \tableline
      D1 &  4.0 & 0.3   & 0.94 \\
      D2 &  4.0 & 0.22  & 0.88 \\
      D3 &  4.0 & 0.31  & 0.80 \\
      D4 &  4.1 & 0.31  & 0.69 \\
      \tableline
      \tablenotetext{a}{From Neill et al. (2006).}
    \end{tabular}
  \end{center}
\end{table}

\clearpage
\begin{table}
  \begin{center}
    \caption{Corrected SNe\,Ia rates per unit mass for D1--4.  \label{TS4NcorrIa} }
    \begin{tabular}{lcccc}
      \tableline\tableline 
      Galaxy            &  $N_{Ia}$  &  $N_{corr,Ia}$     & Stellar Mass                  & SN\,Ia Rate                      \\
      Set              &            &  ($\rm SNe \ y^{-1}$)    & ($\rm  \times 10^{14} \ M_{\odot}$)  & ($\rm \times 10^{-14} \ SNe \ y^{-1} \ M_{\odot}^{-1}$)  \\
      \tableline
      Clusters, 0.4 Mpc  & 2   & $2.8^{+6.4}_{-1.6}$     & 0.33 & $8.3^{+19.4}_{-4.9}$  \\
      Clusters, 0.8 Mpc  & 3   & $4.1^{+6.7}_{-2.1}$     & 0.48 & $8.5^{+14.0}_{-4.3}$  \\
      Clusters, 1.5 Mpc  & 6   & $8.1^{+7.3}_{-3.1}$     & 0.78 & $10.4^{+9.4}_{-4.0}$  \\
      Whole Field        & 109 & $169.1^{+19.8}_{-16.3}$ & 11.4 & $14.8^{+1.7}_{-1.4}$  \\
      \tableline
      \multicolumn{5}{l}{For early-type galaxies only: } \\
      Clusters, 0.4 Mpc  & 2   & $2.8^{ +6.4}_{-1.6}$    & 0.29 & $9.6^{+22.4}_{-5.6}$   \\ 
      Clusters, 0.8 Mpc  & 2   & $2.8^{ +6.4}_{-1.6}$    & 0.38 & $7.2^{+16.8}_{-4.2}$   \\ 
      Clusters, 1.5 Mpc  & 2   & $2.8^{ +6.4}_{-1.6}$    & 0.56 & $5.0^{+11.6}_{-2.9}$   \\ 
      Whole Field        & 23  & $35.1^{+11.0}_{-7.2}$   & 6.23 & $5.6^{+1.8}_{-1.2}$    \\ 
      \tableline
      \multicolumn{5}{l}{For the brightest early-type galaxies only: } \\
      Clusters, 0.4 Mpc  & 1   & $1.3^{+6.1}_{-1.1} $  & 0.13   & $9.8^{+45.9}_{-8.1}$   \\ 
      Clusters, 0.8 Mpc  & 1   & $1.3^{+6.1}_{-1.1} $  & 0.15   & $8.7^{+40.9}_{-7.2}$   \\ 
      Clusters, 1.5 Mpc  & 1   & $1.3^{+6.1}_{-1.1} $  & 0.19   & $7.0^{+33.0}_{-5.8}$   \\ 
      Whole Field        & 2   & $2.6^{+6.3}_{-1.7} $  & 1.02   & $2.5^{+6.2}_{-1.6} $   \\ 
      \tableline
    \end{tabular}
  \end{center}
\end{table}

\clearpage

% FIGURES:
 
%% SIGNIFICANCE SECTION

%% OLSEN SECTION

\end{document}